\documentclass[aps,pra,twocolumn,preprintnumbers,showpacs,superscriptaddress,amsmath,amssymb]{revtex4}
\usepackage{dcolumn}
\usepackage{bm}
\usepackage{amssymb}
\usepackage[german, english]{babel}
\usepackage{graphicx}
\usepackage{color}
\usepackage{booktabs}
\usepackage{dcolumn}
\usepackage[letterpaper,total={7in,9.5in},top=0.75in,left=0.75in]{geometry}

\begin{document}

\title{Quantum degenerate mixtures of strontium and rubidium atoms}

\author{Benjamin Pasquiou}
\affiliation{Institut f\"ur Quantenoptik und Quanteninformation (IQOQI), \"Osterreichische Akademie der Wissenschaften, 6020 Innsbruck, Austria}
\author{Alex Bayerle}
\affiliation{Institut f\"ur Quantenoptik und Quanteninformation (IQOQI), \"Osterreichische Akademie der Wissenschaften, 6020 Innsbruck, Austria}
\affiliation{Institut f\"ur Experimentalphysik und Zentrum f\"ur Quantenphysik, Universit\"at Innsbruck, 6020 Innsbruck, Austria}
\author{Slava Tzanova}
\affiliation{Institut f\"ur Quantenoptik und Quanteninformation (IQOQI), \"Osterreichische Akademie der Wissenschaften, 6020 Innsbruck, Austria}
\affiliation{Institut f\"ur Experimentalphysik und Zentrum f\"ur Quantenphysik, Universit\"at Innsbruck, 6020 Innsbruck, Austria}
\author{Simon Stellmer}
\affiliation{Institut f\"ur Quantenoptik und Quanteninformation (IQOQI), \"Osterreichische Akademie der Wissenschaften, 6020 Innsbruck, Austria}
\author{Jacek Szczepkowski}
\affiliation{Institut f\"ur Quantenoptik und Quanteninformation (IQOQI), \"Osterreichische Akademie der Wissenschaften, 6020 Innsbruck, Austria}
\affiliation{Institute of Physics, Polish Academy of Sciences, 02-668 Warsaw, Poland}
\author{Mark Parigger}
\affiliation{Institut f\"ur Quantenoptik und Quanteninformation (IQOQI), \"Osterreichische Akademie der Wissenschaften, 6020 Innsbruck, Austria}
\affiliation{Institut f\"ur Experimentalphysik und Zentrum f\"ur Quantenphysik, Universit\"at Innsbruck, 6020 Innsbruck, Austria}
\author{Rudolf Grimm}
\affiliation{Institut f\"ur Quantenoptik und Quanteninformation (IQOQI), \"Osterreichische Akademie der Wissenschaften, 6020 Innsbruck, Austria}
\affiliation{Institut f\"ur Experimentalphysik und Zentrum f\"ur Quantenphysik, Universit\"at Innsbruck, 6020 Innsbruck, Austria}
\author{Florian Schreck}
\affiliation{Institut f\"ur Quantenoptik und Quanteninformation (IQOQI), \"Osterreichische Akademie der Wissenschaften, 6020 Innsbruck, Austria}

\date{\today}

\pacs{67.85.-d, 37.10.De}



\begin{abstract}
We report on the realization of quantum degenerate gas mixtures of the alkaline-earth element strontium with the alkali element rubidium. A key ingredient of our scheme is sympathetic cooling of Rb by Sr atoms that are continuously laser cooled on a narrow linewidth transition. This versatile technique allows us to produce ultracold gas mixtures with a phase-space density of up to 0.06 for both elements. By further evaporative cooling we create double Bose-Einstein condensates of $^{87}$Rb with either $^{88}$Sr or $^{84}$Sr, reaching more than $10^5$ condensed atoms per element for the $^{84}$Sr-$^{87}$Rb mixture. These quantum gas mixtures constitute an important step towards the production of a quantum gas of polar, open-shell RbSr molecules.
\end{abstract}

\maketitle

\section{Introduction}
\label{sec:Introduction}

Quantum degenerate gas mixtures of different chemical elements have opened up important new areas for the study of interacting quantum systems. The possibility to apply species-specific optical potentials \cite{LeBlanc2007sso} is a versatile tool and has for example been used to study the exchange of entropy between two gases \cite{Catani2009eei} or mixed-dimensional systems \cite{Lamporesi2010sim}. The mass difference between the constituents of the mixture can lead to new few- and many-body phenomena \cite{Bloch2008mbp,Cugliandolo2010book}, such as novel trimer states \cite{Kartavtsev2007let,Levinsen2009ads,Petrov2010fbp} or crystalline quantum phases \cite{Petrov2007cpo}. Quantum gas mixtures of two elements have also attracted a great deal of attention because they are an ideal starting point for the coherent production of heteronuclear ground-state molecules, which can have large electric dipole moments \cite{Ni2008HighPSDPolMol,Krems2009book,ChemRev2012siu}. The dipole interaction can dominate the behavior of a quantum gas of these molecules and lead to intriguing many-body phenomena \cite{Baranov2012RevDipQGas,Gorshkov2011tsa,Zhou2011lra,Manmana2013tpi}. Ultracold polar molecules also provide insights into chemistry at the quantum level and have the potential to be used as sensitive probes for variations of fundamental constants or as the basis of quantum computation schemes \cite{Krems2009book,Ospelkaus2010QChemReact}.

Most experimentally investigated quantum gas mixtures of two elements consist of two alkali metals \cite{Roati2002fbq,Hadzibabic2002tsm,Taglieber2008qdt,Lercher2011poa,McCarron2011dsb,Wu2011sii,Park2012qdb}. Advances in producing quantum degenerate samples of Yb \cite{Takasu2003YbBEC} and alkaline-earth elements \cite{Kraft2009CaBEC,Stellmer2009bec,MartinezDeEscobar2009SrBEC} have led to efforts towards mixtures containing these elements and recently quantum degenerate Yb-Li mixtures were obtained \cite{Hara2011qdm,Hansen2011YbLiQDeg}. A driving force behind these efforts is the interest in quantum gases of polar molecules beyond alkali dimers, such as RbYb \cite{Munchow2012PAYbRb}, LiYb \cite{Hara2011qdm,Hansen2011YbLiQDeg}, or RbSr \cite{Aoki2013pli}. Contrary to alkali dimers, these open-shell molecules possess an unpaired electron, which provides them with a rich spin structure and a magnetic dipole moment. This property will enable new ways to design and control few- and many-body systems and could prove very useful to implement lattice-spin models \cite{Micheli2006ToolboxPolMol}, to suppress inelastic collisions \cite{Stuhl2012QChemReact}, to imprint geometrical phases \cite{Wallis2009cii}, or to study collective spin excitations \cite{PerezRios2010efc}.

In this Article, we present the realization of quantum gas mixtures composed of the alkali metal $^{87}$Rb and the alkaline-earth metal $^{88}$Sr or $^{84}$Sr. An essential ingredient of our experimental strategy is the use of sympathetic laser cooling \cite{Schreck2001sco,Mudrich2002scw,Ferrari2006cos} on a narrow linewidth transition, which allows us to reach a high phase-space density (PSD) for both elements before evaporative cooling. Strontium atoms are laser cooled on the narrow $^1$S$_0$ $\rightarrow$ $^3$P$_1$ intercombination line and act as a refrigerant for Rb confined in an optical dipole trap. During this sympathetic laser cooling stage, the PSD of Rb increases by a factor of more than 200, with only a 20\% reduction of the Rb atom number. In less than 400\,ms the PSD of both elements can reach 0.06. These very favorable conditions allow us to efficiently reach quantum degeneracy for both species by evaporative cooling and to create BECs with more than $10^5$ atoms per element. The ease of producing large quantum degenerate samples enabled by sympathetic narrow-line laser cooling, together with the large electric dipole moment of RbSr ground-state molecules of 1.5\,Debye \cite{Guerout2010gso}, make Sr-Rb quantum gas mixtures an ideal stepping stone towards the exploration of dipolar physics with open-shell molecules.

The organization of this paper is as follows. In Sec.~\ref{sec:Overview}, we present an overview of our scheme. Section \ref{sec:PreparationStage} describes the loading of a cloud of Rb into an optical dipole trap, followed by the loading of Sr atoms into a narrow-line magneto-optical trap (MOT). Section \ref{sec:SympatheticLaserCoolingStage} focuses on our sympathetic narrow-line laser cooling scheme. In Sec.~\ref{sec:EvaporationStage} we describe the final evaporative cooling stage and the production of two different quantum degenerate mixtures, a $^{88}$Sr-$^{87}$Rb and a $^{84}$Sr-$^{87}$Rb double BEC.

\section{Overview of the experimental strategy}
\label{sec:Overview}

To reach quantum degeneracy, most ultracold atom experiments rely on laser cooling followed by evaporative cooling. The latter process intrinsically leads to a loss of atoms. To minimize this loss, it is beneficial to develop laser cooling methods that are able to reach high PSDs. The main atomic transition for laser cooling of alkalis, such as Rb, is broad, on the order of several MHz. To achieve high PSDs by laser cooling of such species, one can apply sub-Doppler cooling techniques, such as polarization gradient cooling in optical molasses \cite{Lett1988SysiphusExp,Dalibard1989Sysiphus,Modugno1999sdl,Landini2011sdl,Fernandes2012sdl,Grier2013les}, velocity selective coherent population trapping \cite{Aspect1988VelocSelect}, Raman cooling \cite{Kasevich1992RamanCooling}, Raman sideband cooling \cite{Hamann1998ResolvRamanSideband}, or narrow-line cooling on transitions to higher electronic states, which have linewidths down to the 100\,kHz range \cite{Duarte2011aop,McKay2011lth}. The best PSD achieved so far by laser cooling of an alkali is to our knowledge 0.03 \cite{Han2000RamanSidebandCs}. As a member of the alkaline-earth family, Sr has a singlet/triplet electronic structure. Its $^1$S$_0$ $\rightarrow$ $^3$P$_1$ intercombination line has a width of only 7.4\,kHz. It has been shown that simple Doppler cooling using this narrow line can reach temperatures as low as 250\,nK \cite{Loftus2004NarrowLineCooling}. A PSD of 0.1 has been reached by transferring Sr clouds from a narrow-line MOT into an optical dipole trap \cite{Ido2000HighPSD,Stellmer2013QDegSr}. Starting from such samples, evaporative cooling can produce BECs that contain 25\% of the initial atoms, much more than the 1\% typical for alkali BEC experiments. It is therefore tempting to transfer those excellent properties to other species, by using Sr as a cooling agent. Here we demonstrate the efficiency of sympathetic cooling of Rb with Sr atoms laser cooled on a narrow line and the formation of dual-species BECs by a consecutive evaporative cooling stage.

\begin{figure}[htp]
\includegraphics[width=\columnwidth]{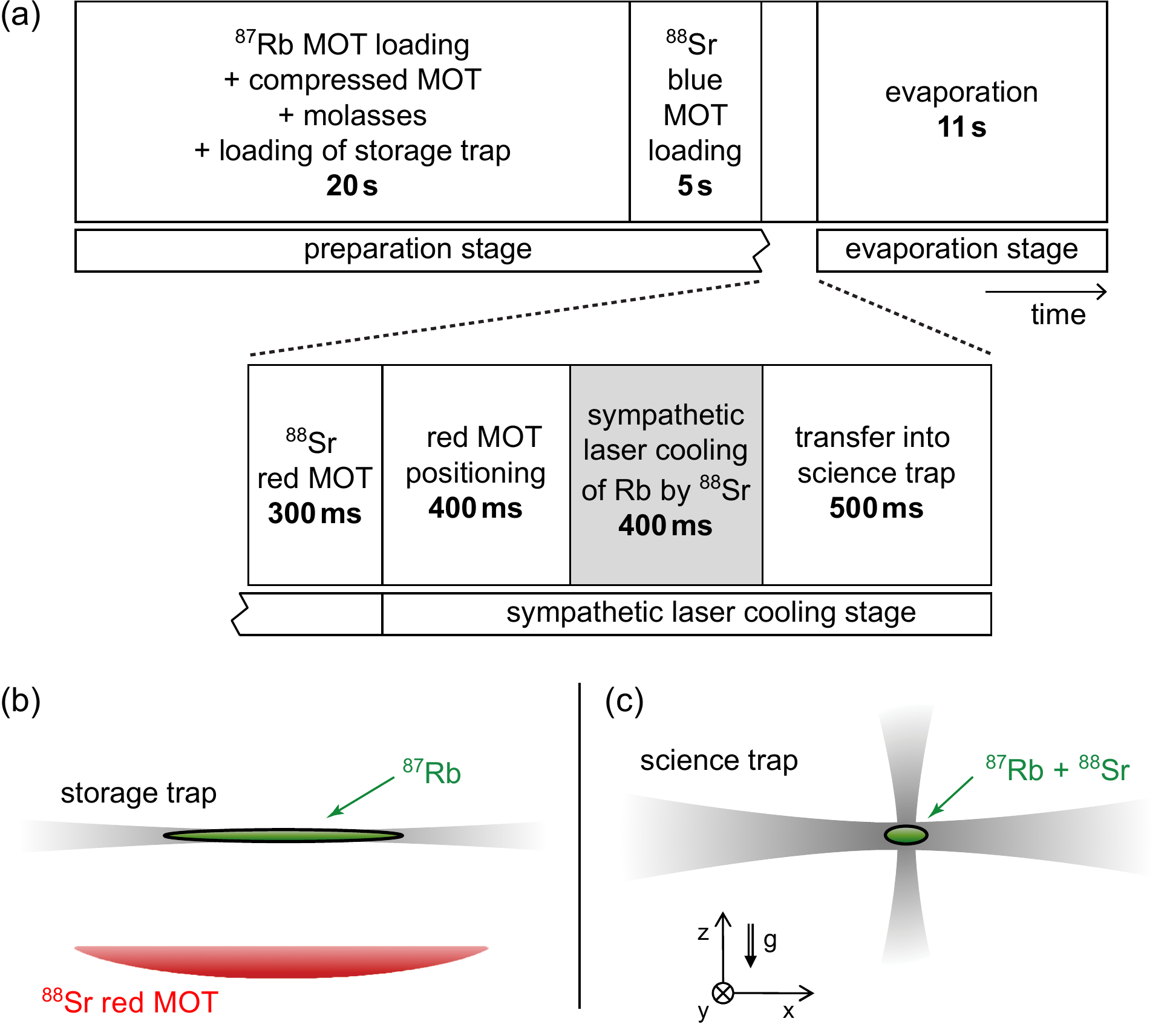}
\caption{\label{fig:Fig1_87Rb88SrBECTimingSequence} (Color online) Timing of the experimental sequence and trap configurations used to produce a $^{88}$Sr-$^{87}$Rb double BEC. (a) Timing sequence. The central sympathetic laser cooling of $^{87}$Rb by $^{88}$Sr shaded in gray is characterized in Fig.~\ref{fig:Fig2_SympatheticLaserCooling}. (b,c) Dipole trap configurations and atomic clouds at the end of the preparation stage (b) and during the evaporation stage (c) (not to scale).}
\end{figure}

Our procedure to achieve quantum degeneracy can be divided into three main stages, see Fig.~\ref{fig:Fig1_87Rb88SrBECTimingSequence}(a). During the first, ``preparation'' stage (Sec.~\ref{sec:PreparationStage}) we use well-established cooling and trapping techniques to prepare ultracold samples of Rb and Sr. We accumulate Rb atoms in a magneto-optical trap and transfer them into a single beam optical dipole trap, henceforth referred to as the ``storage'' trap. After having stored Rb, we operate a ``blue'' Sr MOT, thereby accumulating metastable Sr atoms in a magnetic trap. We then optically pump Sr back to the ground state and capture the atoms in a ``red'', narrow-line MOT, see Fig.~\ref{fig:Fig1_87Rb88SrBECTimingSequence}(b). During the second, ``sympathetic laser cooling'' stage (Sec.~\ref{sec:SympatheticLaserCoolingStage}) we sympathetically cool the Rb cloud by $^{88}$Sr atoms that are continuously laser cooled to a few $\mu$K on a narrow linewidth transition. We then transfer Rb and either $^{88}$Sr or $^{84}$Sr into a second, ``science'' dipole trap, see Fig.~\ref{fig:Fig1_87Rb88SrBECTimingSequence}(c). During the last, ``evaporation'' stage (Sec.~\ref{sec:EvaporationStage}) we perform evaporative cooling, where Sr sympathetically cools Rb until quantum degeneracy is reached for both elements.

\section{Preparation of an ultracold sample of rubidium and strontium}
\label{sec:PreparationStage}

In this Section, we describe our experimental setup and the preparation of an ultracold mixture consisting of Rb contained in the storage trap and Sr stored in a narrow-line MOT, see Fig.~\ref{fig:Fig1_87Rb88SrBECTimingSequence}(b).

Our experimental setup is based on our Sr BEC apparatus, which has been described in detail in \cite{Stellmer2013QDegSr,StellmerPhD}. The basic principle of the apparatus is to capture a Zeeman slowed atomic beam in a MOT and to cool the gas to quantum degeneracy by evaporation out of a dipole trap. Here we will focus on the upgrades carried out to also trap and cool Rb with the apparatus. Copropagating atomic beams of each element are produced by two independent ovens, heated to 550\,$^{\circ}$C for Sr and 200\,$^{\circ}$C for Rb \cite{WillePhD,StellmerPhD}. The same Zeeman slower (ZS) is used to slow both beams. Since the magnetic fields required to slow Rb and Sr differ significantly, we perform a time-sequential loading scheme. The Rb ZS laser beam is superposed with the Sr ZS beam by a dichroic mirror, and uses 14\,mW of light detuned by $-165\,$MHz from the $^2$S$_{1/2}$-{$^2$P$_{3/2}$}, $|F=2\rangle \rightarrow |F'=3\rangle$ transition. A ``repumping'' beam with 4.5\,mW of power, addressing the $|F=1\rangle \rightarrow |F'=1\rangle$ transition is overlapped with the ZS beam. All three laser beams have a waist of about 8\,mm at the MOT position. Slowed atoms are captured by a 3D MOT. Producing an ultracold mixture of Rb and Sr requires the use of MOTs of three wavelengths, one for Rb and two for Sr. The Rb MOT uses 780-nm light, the blue Sr MOT is operated on the broad $^1$S$_0$ $\rightarrow$ $^1$P$_1$ transition at 461\,nm, and the red MOT on the $^1$S$_0$ $\rightarrow$ $^3$P$_1$ intercombination line at 689\,nm. MOT beams of the three wavelengths are overlapped by dichroic mirrors on each of the three retro-reflected MOT beam paths. The Rb MOT beams have a waist of 9.8\,mm, a power of 18\,mW (25\,mW) in the horizontal (vertical) direction, and are detuned by $-16\,$MHz from the $^2$S$_{1/2}$-{$^2$P$_{3/2}$}, $|F=2\rangle \rightarrow |F'=3\rangle$ transition. A repumping beam with a waist of 6\,mm and a peak intensity of 130\,$\mu$W/cm$^2$, on resonance with the $|F=1\rangle \rightarrow |F'=2\rangle$ transition, is shone onto the MOT. The quadrupole magnetic field of the Rb MOT has a gradient of 12\,G/cm along the vertically oriented coil axis.

The experimental sequence starts by operating the Rb MOT during 20\,s to accumulate a cloud of $2\times10^7$ atoms at a temperature of 175\,$\mu$K. We then compress the cloud in 140\,ms by raising the gradient of the quadrupole field to 50\,G/cm and increasing the MOT laser detuning to $-30\,$MHz. After compression the $1/e$ cloud radius is $\sim250\,\mu$m. To decrease the temperature we use polarization gradient cooling in an optical molasses. After switching off the magnetic field, the MOT laser detuning is set to $-110\,$MHz and the beam power is halved. After 3\,ms of molasses, we obtain $1.8\times10^7$ atoms cooled to 15\,$\mu$K with a peak density of $3\times10^{10}$\,cm$^{-3}$.

We then transfer the Rb atoms into the storage trap. This trap consists of a horizontal beam with a waist of 40\,$\mu$m propagating at a small angle to the x-direction, see Fig.~\ref{fig:Fig1_87Rb88SrBECTimingSequence}(b). The beam is derived from a {100-W} multimode fiber laser operating at a wavelength of 1070\,nm (YLR-100-LP-AC-Y12 from IPG). It is linearly polarized in the vertical direction to minimize the light shift induced on the red Sr laser cooling transition \cite{Stellmer2013QDegSr}. Initially we use a power of 14\,W for the storage trap, which results in a potential depth of $k_B\times830\,\mu$K and trap frequencies of $f_{\rm rad} = 2.2$\,kHz and $f_{\rm ax} = 13$\,Hz in the radial and axial directions respectively. To improve loading of the storage trap, we toggle the repumping beam to a path were a wire is imaged onto the trap region, creating a dark spot. The repumping beam power is reduced to a peak intensity of 8\,$\mu$W/cm$^2$ and in 500\,ms we transfer 10\% of the molasses atoms into the trap, pumping them at the same time into the $F=1$ manifold. Up to $1.7\times10^6$ atoms are stored in the storage trap at a density of $2\times10^{13}$\,cm$^{-3}$ and a temperature of $\sim20\,\mu$K.

Having stored Rb, we now capture Sr atoms. The loading and cooling of Sr is done in a manner similar to our previous work \cite{Stellmer2009bec,Stellmer2013QDegSr}. We operate a blue MOT on the broad transition at 461\,nm, which has a leak towards the metastable $^3$P$_2$ state. We accumulate $^3$P$_2$ atoms in the magnetic trap formed by the quadrupole field of the MOT. We typically load this reservoir in a few seconds with several million $^{88}$Sr atoms. For the data presented in Sec.~\ref{sec:PreparationStage} and \ref{sec:SympatheticLaserCoolingStage} we load for 0.5\,s. The atoms are subsequently optically pumped back to the ground state using a flash of light on the {$^3$P$_2$ $\rightarrow$ {$^3$D$_2$}} transition at 497\,nm. The atoms are captured by a red MOT operating on the $^1$S$_0$ $\rightarrow$ $^3$P$_1$ intercombination transition and using a magnetic field gradient of 1.8\,G/cm. The narrow, 7.4\,kHz-linewidth intercombination transition allows us to cool Sr to less than 1\,$\mu$K while keeping millions of atoms. The atoms settle in the lower part of an ellipsoid of constant magnetic field magnitude, see Fig.~\ref{fig:Fig1_87Rb88SrBECTimingSequence}(b). The size and position of this cloud can be influenced by the magnetic field and the detuning of the MOT light. Varying these parameters facilitates the transfer of Sr into the optical dipole trap.

At the end of the preparation stage, we obtain $7\times10^6$ $^{88}$Sr atoms at a temperature of $2.5\,\mu$K in a red MOT. The storage trap contains $1.6\times10^6$ Rb atoms at a temperature of $30\,\mu$K and a phase-space density of $2.5\times10^{-4}$, which is slightly worse than before loading the Sr atoms. The $1/e$ lifetime of the Rb cloud in the storage trap in presence of the blue $^{88}$Sr MOT is 5.0(5)\,s. To detect the influence of Sr atoms on the Rb cloud in the storage trap, we perform the same experimental sequence, but shutter the atomic beam off after loading the Rb MOT. Under these conditions, the lifetime of the Rb sample in the storage trap is 30(1)\,s. This value does not change if also the Sr laser cooling beams are off.

\section{Sympathetic narrow-line laser cooling}
\label{sec:SympatheticLaserCoolingStage}

The following, sympathetic laser cooling stage is crucial in our approach to obtain a Sr-Rb double BEC. We make use of laser cooled Sr to further cool Rb and increase its PSD. To this aim, we overlap the cloud of $^{88}$Sr atoms with the focus of the storage trap. This overlap is achieved by moving the magnetic field center upwards, while keeping the MOT laser frequency red detuned by 200\,kHz from resonance. At this point, the Sr density is so low that the influence on the Rb cloud is negligible on a timescale of 500\,ms. To increase the density of the Sr cloud, the magnetic field together with the MOT detuning and intensity are changed over 200\,ms such that the cloud is compressed while remaining at the same position.

\begin{figure}[htp]
\includegraphics[width=\columnwidth]{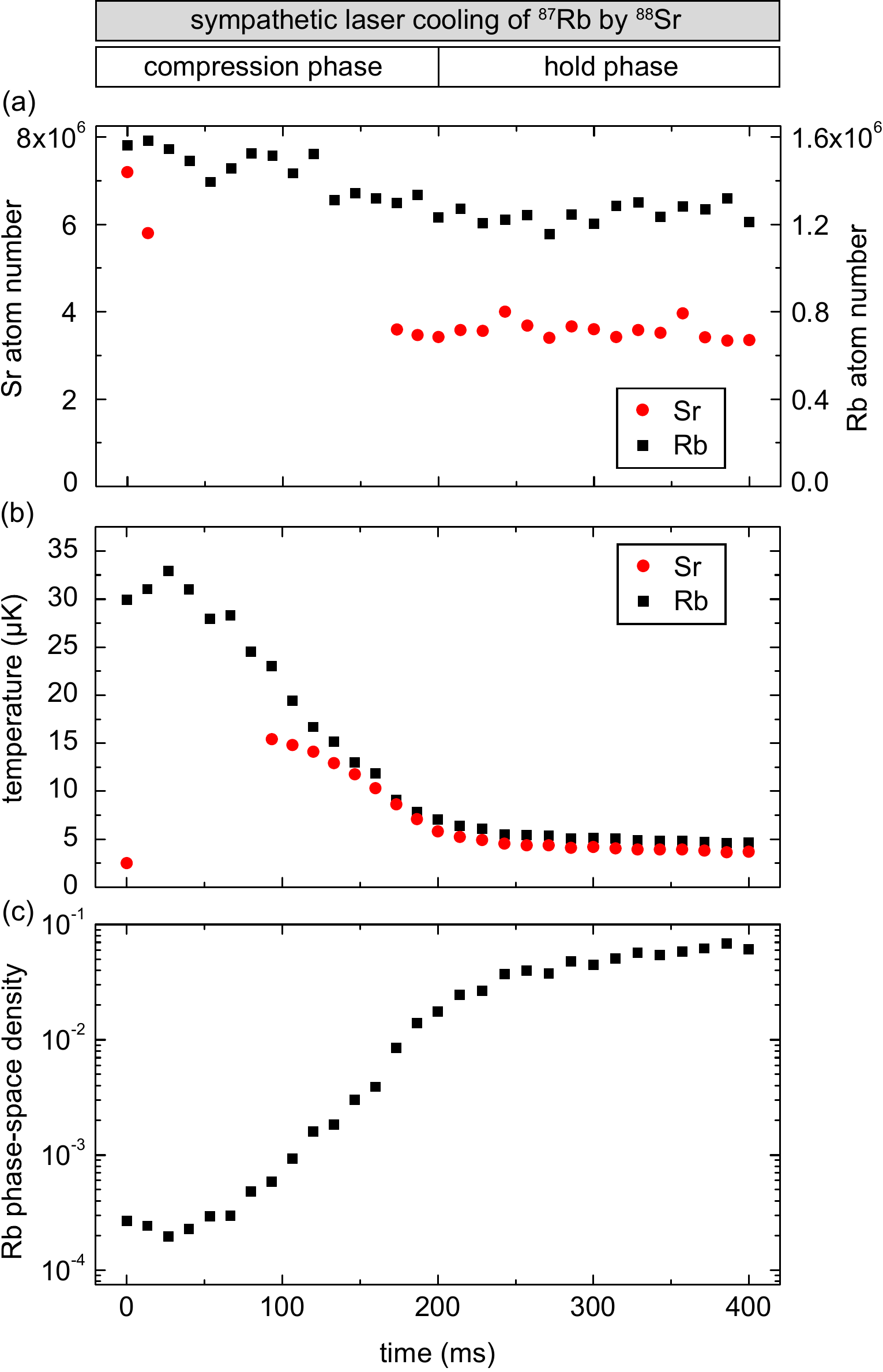}
\caption{\label{fig:Fig2_SympatheticLaserCooling} (Color online) Sympathetic laser cooling of $^{87}$Rb by $^{88}$Sr. During the compression phase the density of the red Sr MOT is increased and the Sr atoms are loaded into the storage dipole trap by changing magnetic field and both the MOT laser detuning and intensity. During the hold phase these parameters are held constant. (a,b) Evolution of the Rb and Sr atom numbers (a) and temperatures (b) as determined from time-of-flight absorption images. During a part of the compression phase the Sr cloud separates into two components, one being heated out of the system, the other remaining trapped and being transferred into the storage trap. We only show the number of trapped Sr atoms and the Sr temperature if we can determine them reliably from bimodal fits to the absorption images. (c) PSD of Rb, calculated from trap parameters and Rb atom number and temperature.}
\end{figure}

The Sr atoms are loaded into the storage trap during this compression phase. The loading process is strongly influenced by the light shift induced on the red MOT transition by this trap. At the center of the trap, the light shift is $\sim+500$\,kHz, which is almost 70 times the linewidth. In order to laser cool Sr atoms in the storage trap we tune the MOT laser frequency $350$\,kHz to the blue of the unshifted atomic transition in a 200\,ms ramp, while reducing the peak beam intensity to 6\,$\mu$W/cm$^2$. The part of the Sr cloud that does not spatially overlap with the dipole trap is thereby expelled from the MOT, leading to a loss of 60\% of the Sr atoms. This loss could be reduced by canceling the light shift of the transition, for example by inducing a light shift on the excited state of the laser cooling transition \cite{Stellmer2013lct}. Since $^{88}$Sr has a high natural abundance, leading to a high atom number in the MOT, we here simply tolerate the loss. The temperature of the Sr cloud in the storage trap increases to 15\,$\mu$K before reducing to below $5\,\mu$K during the compression phase and a subsequent 200\,ms hold phase. This temperature increase results from MOT dynamics in presence of the light shift induced by the dipole trap and would also occur in absence of Rb atoms. The Rb cloud thermalizes with laser cooled Sr by elastic collisions. During this process, the Rb PSD increases dramatically, by a factor of more than 200, reaching 0.062(6), while only 20(5)\% of the Rb atoms are lost. From the thermalization behavior we can deduce a minimum absolute value of the interspecies scattering length of 30\,$a_0$ between $^{87}$Rb and $^{88}$Sr \cite{Monroe1993moc}. After the hold phase the Sr laser cooling beams are switched off. We observe that complete thermal equilibrium between Rb and Sr can only be reached in absence of the cooling light.

To prepare for the creation of a quantum degenerate sample, we transfer both elements into the science dipole trap, which has been described in detail in Ref.~\cite{Stellmer2013QDegSr} and will be especially important for the creation of a large $^{84}$Sr-$^{87}$Rb double BEC (see Sec.~\ref{subsec:Rb87Sr84BEC}). This crossed-beam dipole trap is composed of an elliptical beam propagating in the x-direction with waists of $w_z=17\,\mu$m and $w_x=300\,\mu$m, crossed by a nearly vertical beam with a waist of 90\,$\mu$m, see Fig.~\ref{fig:Fig1_87Rb88SrBECTimingSequence}(c). The center of this dipole trap is overlapped with the center of the storage trap, and the horizontal beams of the two traps intersect at an angle of $17^{\circ}$. Individual laser sources are used for each science trap beam (5-W, 1065-nm multimode fiber lasers, YLD-5-LP from IPG). At the wavelength of 1065\,nm the polarizability of Rb is 2.9 times the polarizability of Sr \cite{Boyd2007hps,Safronova2006PolarizabilityAlkali}. The ratio of trap depths is even higher because of gravitational sagging. Initially the science trap has a depth of $k_B\times25\,\mu$K for Rb and $k_B\times7\,\mu$K for Sr. The reduction of the Rb cloud volume and temperature by sympathetic laser cooling allows us to transfer Rb from the storage trap into this much shallower trap in 500\,ms with a nearly perfect transfer efficiency (more than 95\% of the atoms). The $^{88}$Sr transfer efficiency is 30\%. After transfer we typically obtain $1.25\times10^6$ Rb atoms and $1.6\times10^6$ Sr atoms at a temperature of 1.2\,$\mu$K. The Rb PSD is 0.5(2) and the Sr PSD is 0.10(3). We attribute the increase in PSD to evaporation of Sr during the transfer process.

\section{Evaporation to alkali/alkaline-earth double BECs}
\label{sec:EvaporationStage}

In this Section we present the creation of quantum degenerate mixtures of $^{87}$Rb with either $^{88}$Sr (Sec.~\ref{subsec:Rb87Sr88BEC}) or $^{84}$Sr (Sec.~\ref{subsec:Rb87Sr84BEC}). These two Sr isotopes have markedly different properties, which we take into account in our experimental strategy. The $^{88}$Sr isotope has a high natural abundance and provides us with an ideal coolant for sympathetic narrow-line laser cooling of $^{87}$Rb. Since the $^{88}$Sr scattering length is negative (-2\,$a_0$), the $^{88}$Sr BEC atom number is limited to a few thousand atoms. By contrast, $^{84}$Sr has a convenient scattering length of +123\,$a_0$, allowing us to create BECs with high atom number. Unfortunately this isotope has a low natural abundance (only 0.56\%), which renders it less favorable for sympathetic laser cooling. To overcome this drawback, we employ both isotopes in the production of a $^{84}$Sr-$^{87}$Rb double BEC, using $^{88}$Sr for sympathetic laser cooling of $^{87}$Rb and $^{84}$Sr for evaporative cooling to quantum degeneracy.

\begin{figure}[htp]
\includegraphics[width=\columnwidth]{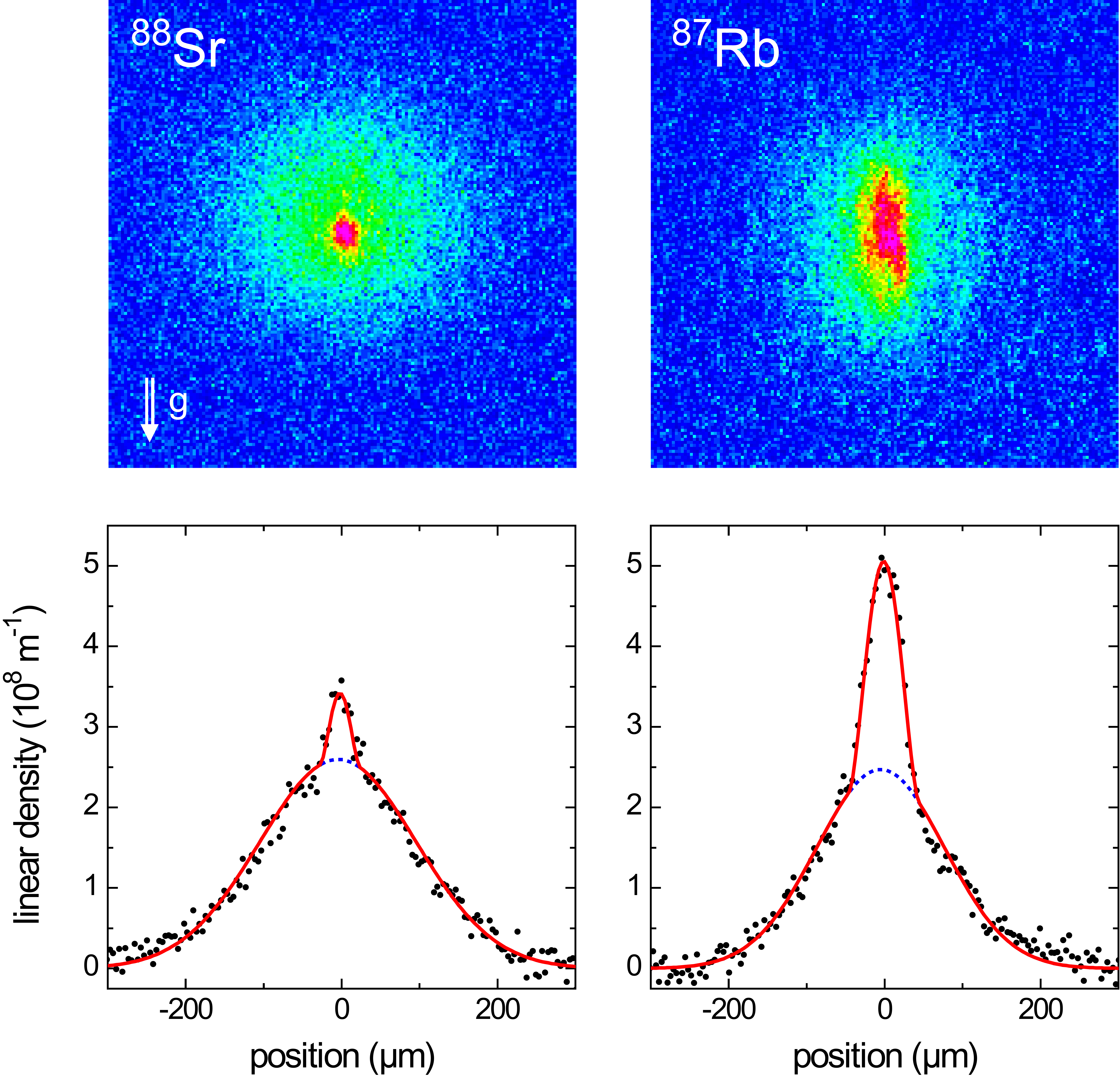}
\caption{\label{fig:Fig3_87Rb88SrBEC} (Color online) $^{88}$Sr-$^{87}$Rb double BEC. The absorption images have been recorded after an expansion time of 26\,ms. The lower panels show density profiles obtained by vertical integration of the absorption images. Bimodal fits consisting of a Gaussian and a Thomas-Fermi distribution are shown as red, solid lines. The Gaussian part of the fit corresponds to the thermal fraction of the cloud and is shown as blue, dotted line.}
\end{figure}

\subsection{$^{88}$Sr-$^{87}$Rb double BEC}
\label{subsec:Rb87Sr88BEC}

To obtain a $^{88}$Sr-$^{87}$Rb double BEC, we prepare as before a $^{88}$Sr-$^{87}$Rb mixture in the science trap. Compared to the previous sections, we here increase the number of ultracold $^{88}$Sr atoms by increasing the Sr MOT loading time to 5\,s. Then we perform forced evaporative cooling by lowering the trap depth exponentially over 11\,s to $0.5\,\mu$K for Sr. As demonstrated in Sec.~\ref{sec:SympatheticLaserCoolingStage}, the interspecies scattering cross section is sufficient for interspecies thermalization. Note that by itself $^{88}$Sr does barely thermalize because of its small scattering length of $-2\,a_0$ \cite{MartinezDeEscobar2008TwoPhotSpectro, Stein2008FourierTransfSpectroSr2} and only the presence of Rb ensures proper thermalization. Since the trap depth is more than three times deeper for Rb than for Sr, Rb is sympathetically cooled by evaporating Sr.

At the end of the evaporation stage, we obtain Rb and Sr BECs with low atom number immersed in thermal clouds (see Fig.~\ref{fig:Fig3_87Rb88SrBEC}). The Sr cloud contains $2.3\times10^3$ condensed atoms and $6.5\times10^4$ thermal atoms. The Rb cloud consists of a mixture of all three $m_F$ states of the $F=1$ manifold, as confirmed by Stern-Gerlach measurements. The spin state distribution can be influenced by applying a magnetic field offset and gradient during the evaporation stage \cite{Chang2001SpinDyn}. The distribution for zero offset and gradient is 38\%, 25\%, and 37\% for the $m_F = -1$, 0, and $+1$ states after evaporation. In total the Rb cloud consists of $1.3\times10^4$ atoms in the spinor BEC and $5\times10^4$ thermal atoms. The temperature of both elements is 190(30)\,nK. The low $^{88}$Sr BEC atom number is expected since the negative scattering length of $^{88}$Sr leads to a collapse of the BEC for higher atom numbers \cite{Dalfovo1999tob,Donley2001doc,Mickelson2010Sr88BEC,Yan2013ccc,Stellmer2013QDegSr}.

\subsection{$^{84}$Sr-$^{87}$Rb double BEC}
\label{subsec:Rb87Sr84BEC}

To obtain quantum degenerate samples with higher atom numbers, we now turn to the $^{84}$Sr isotope. This isotope has a scattering length of $+123\,a_0$ \cite{MartinezDeEscobar2008TwoPhotSpectro, Stein2008FourierTransfSpectroSr2}, which is well-suited for evaporative cooling. Despite the small $^{84}$Sr natural abundance of only 0.56\% the production of large $^{84}$Sr BECs with 10$^7$ atoms has been demonstrated \cite{Stellmer2013QDegSr}. The low abundance can be compensated for by a longer blue MOT duration compared to the one used for the highly abundant $^{88}$Sr isotope, leading to nearly the same atom number accumulated in the metastable state reservoir. By simply replacing $^{88}$Sr with $^{84}$Sr and operating the blue MOT for 20\,s, we can prepare a $^{84}$Sr-$^{87}$Rb double BEC with essentially the same scheme as the one used for the production of a $^{88}$Sr-$^{87}$Rb double BEC. Nonetheless we have developed an improved strategy, which requires less time, makes optimal use of the precious $^{84}$Sr atoms, and leads to much larger numbers of condensed atoms. Because of its high natural abundance, we use $^{88}$Sr as the refrigerant for sympathetic laser cooling of Rb. We then use $^{84}$Sr during the evaporation stage to obtain a $^{84}$Sr-$^{87}$Rb BEC.

\begin{figure}[htp]
\includegraphics[width=\columnwidth]{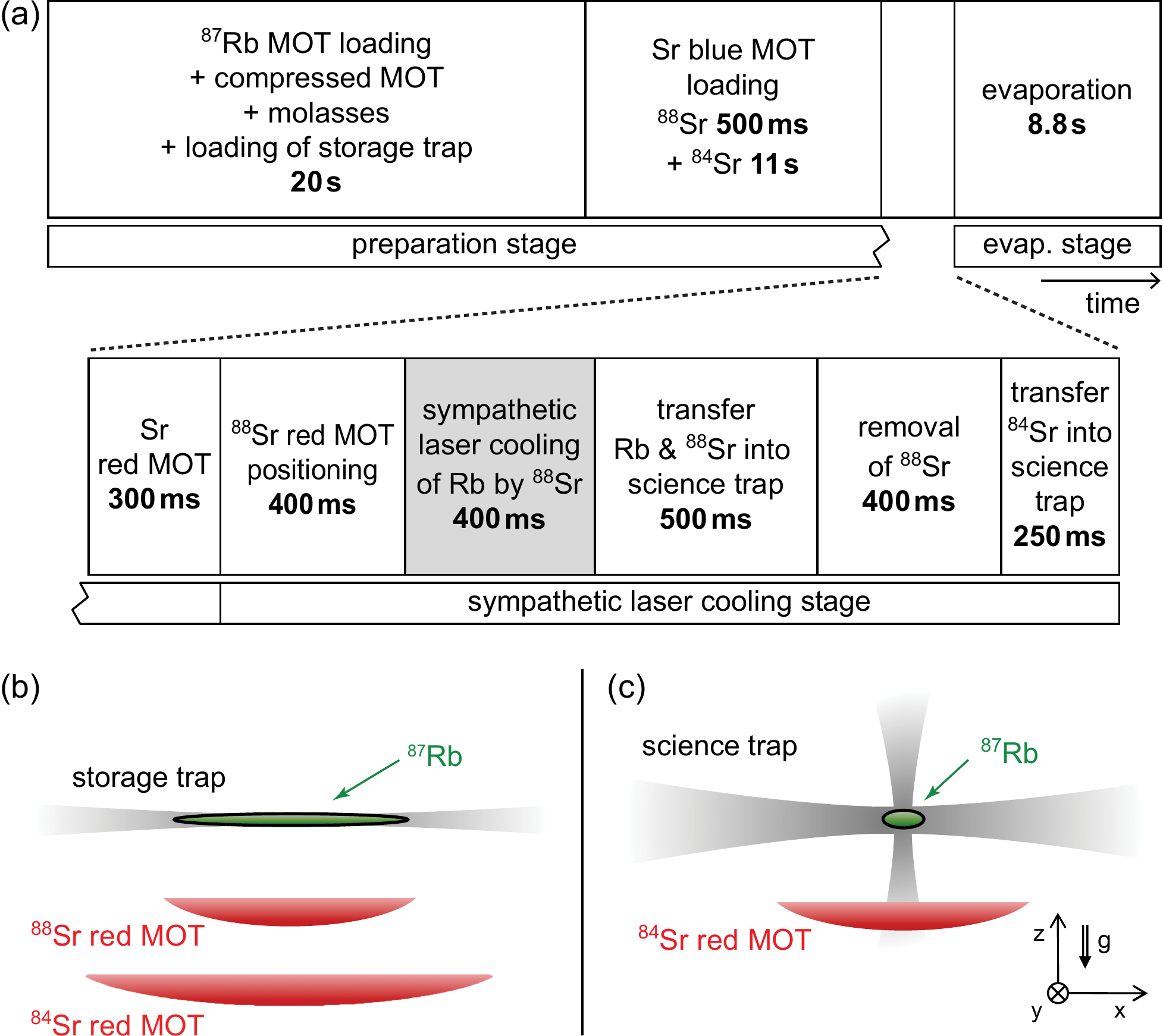}
\caption{\label{fig:Fig4_87Rb84SrBECBECTimingSequence}  (Color online) Timing of the experimental sequence and trap configurations used to produce a $^{84}$Sr-$^{87}$Rb double BEC. (a) Timing sequence. $^{88}$Sr is used for sympathetic laser cooling of Rb and discarded afterwards. $^{84}$Sr, stored in a red MOT during sympathetic laser cooling, is transferred into the science dipole trap before evaporative cooling. (b,c) Dipole trap configurations and atomic clouds at the end of the preparation stage (b) and after removing $^{88}$Sr (c)  (not to scale).}
\end{figure}

We adapt the scheme of our experiment to this new strategy, see Fig.~\ref{fig:Fig4_87Rb84SrBECBECTimingSequence}(a). During the Sr blue MOT stage we load both Sr isotopes, $^{88}$Sr and $^{84}$Sr. This double-isotope loading is achieved by accumulating one isotope after the other in the metastable state reservoir. In between, the frequency of the 461-nm cooling laser source is changed by the isotope shift \cite{Poli2005SrMix,Stellmer2013QDegSr}. We first accumulate $^{88}$Sr for 500\,ms and then add  $^{84}$Sr during 10\,s. The lifetime of Rb atoms in the storage trap in presence of the $^{84}$Sr blue MOT is 23(1)\,s. Afterwards both isotopes are optically repumped into the electronic ground state. The two isotopic clouds are captured simultaneously by two narrow-line red MOTs, which contain $4\times10^6$ atoms of $^{88}$Sr and $9\times10^6$ atoms of $^{84}$Sr respectively, see Fig.~\ref{fig:Fig4_87Rb84SrBECBECTimingSequence}(b) \cite{Poli2005SrMix,Stellmer2013QDegSr}. The different isotopes are addressed independently by using two frequencies for the red MOT laser beams, separated by the isotope shift. Changing one of the MOT laser frequencies, vertically displaces the corresponding isotopic cloud and changes its radial size.

We proceed as described in Sec.~\ref{sec:SympatheticLaserCoolingStage} with the sympathetic laser cooling of Rb by $^{88}$Sr and the transfer of the $^{88}$Sr-$^{87}$Rb mixture into the science trap. In the meantime we keep the $^{84}$Sr cloud about 0.7\,mm below the center of the dipole traps. At this position the red MOT has a radius of about 1\,mm and a low density, which reduces light-assisted collisions. The lifetime of the MOT is independent of the various operations on the other species and we lose only 10\% of the $^{84}$Sr atoms during the sympathetic laser cooling stage. Reciprocally, the presence of the $^{84}$Sr red MOT does not affect the sympathetic laser cooling of Rb.

It is advantageous to expel the refrigerant $^{88}$Sr after it has fulfilled its role. Without the removal of $^{88}$Sr, the large scattering length between $^{84}$Sr and $^{88}$Sr of about $1700\,a_0$ \cite{MartinezDeEscobar2008TwoPhotSpectro, Stein2008FourierTransfSpectroSr2} would lead to strong three-body loss as soon as $^{84}$Sr is loaded into the science trap. We expel $^{88}$Sr by adiabatically lowering the science trap depth in 300\,ms. Including gravitational sagging the trap is about ten times shallower for Sr than for Rb at the end of the ramp, leading to a removal of all $^{88}$Sr atoms without affecting the Rb atom number. We then raise the trap back to its former depth in 100\,ms.

\begin{figure}[htp]
\includegraphics[width=\columnwidth]{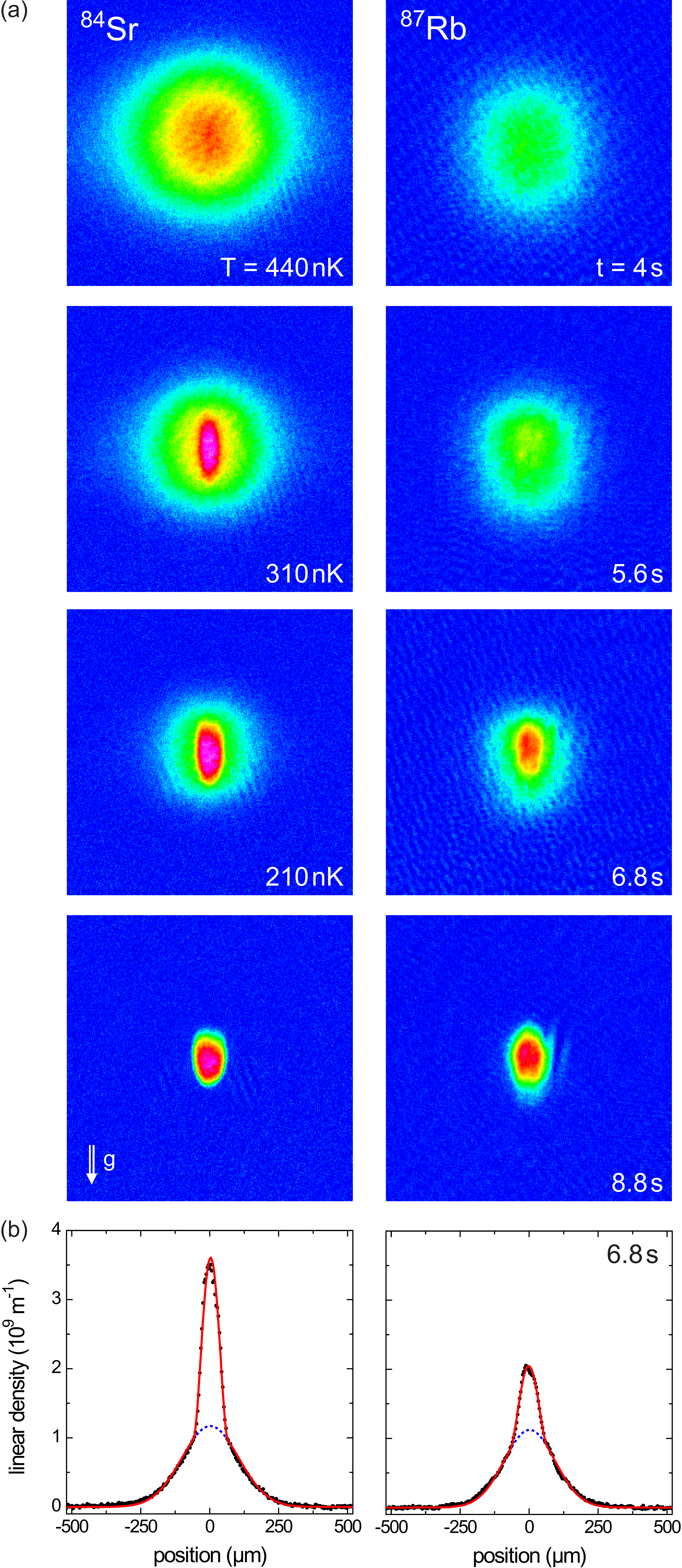}
\caption{\label{fig:Fig5_87Rb84SrBEC} (Color online) Formation of a $^{84}$Sr-$^{87}$Rb double BEC while Sr is evaporated, sympathetically cooling Rb. (a) Absorption images of the Rb and Sr clouds at time $t$ of the evaporative cooling ramp recorded after 24\,ms of expansion. Strontium condenses at $t=4.3\,s$ and Rb at $t=6\,s$. The temperature $T$ of Sr is given. (b) Density profiles obtained from the absorption images at $t=6.8\,$s by integration along the vertical direction. The solid, red lines are bimodal fits to the data by a Gaussian plus a Thomas Fermi distribution. The blue, dotted line shows the Gaussian part of the fit corresponding to the thermal part of the cloud.}
\end{figure}

At this point we load $^{84}$Sr into the science trap, see Fig.~\ref{fig:Fig4_87Rb84SrBECBECTimingSequence}(c). We shift the red MOT upwards by changing the cooling laser frequency until the MOT overlaps with the science trap. Now 70\% of the $^{84}$Sr atoms are loaded into the dipole trap, compared to 40\% when transferring into the storage trap as in Sec.~\ref{sec:SympatheticLaserCoolingStage}. The reasons for this increased transfer efficiency are that the much shallower science trap induces a negligible light shift on the red Sr laser cooling transition and that the horizontally extended science trap is well adapted to the pancake shape of the red MOT. The $^{84}$Sr cloud is slightly colder than the Rb sample, which again leads to sympathetic laser cooling of Rb. When choosing the final red MOT parameters, we have to compromise between attainable temperature and remaining Rb and Sr atom numbers \cite{Loftus2004NarrowLineCooling,Ferrari2006cos}. We chose a temperature of $1\,\mu$K, for which we obtain $4.0(1)\times10^6$ Sr atoms and $5.2(1)\times10^5$ Rb atoms, both elements at a PSD of 0.4(1). At this point the cooling laser beams are switched off.

Starting with these excellent conditions, we perform evaporative cooling by lowering the science trap depth exponentially over 8.8\,s to $k_B\times150$\,nK for Sr. Strontium is evaporated, sympathetically cooling Rb, and a double BEC is formed (see Fig.~\ref{fig:Fig5_87Rb84SrBEC}). At the end of evaporation we obtain a pure $^{84}$Sr BEC of $2.3\times10^5$ atoms and $1.3\times10^5$ quantum degenerate Rb atoms accompanied by $6.5\times10^{4}$ thermal Rb atoms at a temperature of $\sim70\,$nK. The Rb cloud again contains a nearly equal mixture of the three $F=1$ $m_F$ states. The trapping frequencies in the x-, y-, and z-direction are at this point 40\,Hz, 37\,Hz, and 190\,Hz for Sr and 67\, Hz, 63\,Hz, and 400\,Hz for Rb. For the lowest values of the science trap depth, the gravitational sagging of the Sr cloud is $\sim3.5\,\mu$m larger than the gravitational sagging of the Rb cloud. For comparison, the Thomas-Fermi radii, calculated neglecting the interspecies mean-field, is 2.5\,$\mu$m for the Sr BEC in the vertical direction and 2\,$\mu$m for the Rb BEC. The two elements are barely overlapping, which reduces interspecies thermalization. The differential gravitational sag between Sr and Rb could be compensated for the magnetic Rb $F=1$, $m_F=+1$ or $m_F=-1$ state by using a magnetic field gradient in the vertical direction, which does not influence the non-magnetic Sr. The peak density of the Sr BEC and of each $m_F$-state component of the Rb BEC is $1.5\times10^{14}$\,cm$^{-3}$. The $1/e$ lifetime of the BECs is about 10\,s. The atom number of the Sr BEC can be increased at the expense of the Rb BEC atom number by reducing the MOT loading time of Rb. On absorption images of $^{84}$Sr-$^{87}$Rb double BECs taken after 24\,ms of expansion, we observe that the position of the Sr BEC is shifting downwards for an increasing Rb BEC atom number. This observation hints at a positive mean-field interaction and therefore a positive interspecies scattering length between $^{87}$Rb and $^{84}$Sr.

\section{Conclusion and outlook}
\label{sec:conclusion}

We have presented the production of $^{88}$Sr-$^{87}$Rb and $^{84}$Sr-$^{87}$Rb double BECs. Crucial to our success are the favorable interaction properties of the two mixtures. For both mixtures we observe efficient thermalization. At the same time the mixtures do not suffer from large inelastic three-body losses. These interaction properties cannot be predicted by \emph{ab-initio} calculation and were completely unknown prior to our work.

A central stage in our scheme is sympathetic narrow-line laser cooling of Rb by Sr. This powerful technique will also be useful to cool other species besides $^{87}$Rb, including fermions, and should work if one of the four Sr isotopes has good interspecies scattering properties with the target species. It might even be possible to sympathetically laser cool the target species to quantum degeneracy without using evaporation \cite{Stellmer2013lct}. This goal will be facilitated by selectively increasing the density of the target species with a species-specific dipole potential \cite{LeBlanc2007sso} or by using a target species of low mass, which leads to a high critical temperature or Fermi temperature for a given density.

Our next goal is the creation of RbSr molecules. We plan to associate atoms to weakly-bound molecules by either magneto-association \cite{Zuchowski2010urm} or stimulated Raman adiabatic passage (STIRAP) \cite{Vitanov2001lip,Stellmer2012Sr2Mol}. These molecules will then be transferred into the ro-vibrational ground-state by STIRAP \cite{Danzl2008qgo,Lang2008utm,Ni2008ahp}. We are currently performing photoassociation spectroscopy of the $^{84,88}$Sr-$^{87}$Rb mixtures, in order to precisely determine the interspecies scattering lengths of all Sr-Rb isotopic combinations, the magnetic field values of Sr-Rb magnetic Feshbach resonances, and STIRAP paths for molecule association and ground-state transfer. The optimal Sr-Rb isotopic mixture for our task will depend on these properties, especially the interspecies scattering length, which determines the miscibility of the Rb and Sr quantum gases.

\begin{acknowledgments}
We thank Piotr \.{Z}uchowski, Olivier Dulieu, Jeremy Hutson, John Bohn, Roman Krems, Kadan Hazzard, Paul Julienne, Mikhail Baranov, and Mikhail Lemeshko for fruitful discussions. We gratefully acknowledge support from the Austrian Ministry of Science and Research (BMWF) and the Austrian Science Fund (FWF) through a START grant under Project No. Y507-N20. As member of the project iSense, we also acknowledge the financial support of the Future and Emerging Technologies (FET) programme within the Seventh Framework Programme for Research of the European Commission, under FET-Open grant No. 250072.
\end{acknowledgments}

\bibliographystyle{apsrev}

\end{document}